\documentclass[journal]{vgtc}

\onlineid{0}
\vgtccategory{System or Tool}

\title{GeoXplain: On-the-Fly Visual Explanations for Weather Foundation Models}

\author{Clemens Walter Koprolin,
Leonardo Trentini,
Benedikt Soja,
Mennatallah El-Assady,
Christina Humer
}

\affiliation{\scriptsize ETH Zurich, Zurich, Switzerland}
\authorfooter{
\item[] E-mail: \{ckoprolin\,$|$\,ltrentini\,$|$\,sojab\,$|$\,melassad\,$|$\,humerc\}@ethz.ch
}

\manuscriptnote{}

\abstract{
Weather and climate foundation models produce high-dimensional forecasts whose learned relationships are difficult to inspect with static plots alone.
GeoXplain is an interactive Python-based visualization toolkit for exploring geospatial attribution maps across climate variables, atmospheric pressure levels, and forecast time.
The toolkit accepts attribution bundles containing attribution grids together with corresponding metadata and renders them in a notebook widget or browser with map and globe modes, linked timelines, pressure-level controls, target annotations, and optional physical-field overlays.
We frame GeoXplain as a model-agnostic earth-system visualization toolkit and present the GeoXplain Aurora Adapter as its first computation backend.
The adapter computes explanations for the Aurora foundation model, either in a local GPU process, through a GPU listener, or through a SLURM-backed listener, while preserving the same Python call site for analysts.
It currently supports gradient saliency, Integrated Gradients, RISE, ViT-CX, multi-frame saliency and Integrated Gradients rollouts, and retrieval of ERA5 overlays. GeoXplain can be installed as a PyPI package with \texttt{pip install geoxplain}. The code is open-source and available at \url{https://github.com/clemenskoprolin/geoxplain}.
}

\keywords{Climate visualization, geospatial visualization, explainable AI, weather foundation models, computational notebooks}

\graphicspath{{figs/}{figures/}{pictures/}{images/}{./}}

\usepackage{booktabs}
\usepackage{amsmath}
\usepackage{amssymb}
\usepackage{mathptmx}
\usepackage{array}
\usepackage{dblfloatfix}
\usepackage{float}

\teaser{
  \centering
  \includegraphics[
    width=0.94\linewidth,
    height=3.7in,
    keepaspectratio
  ]{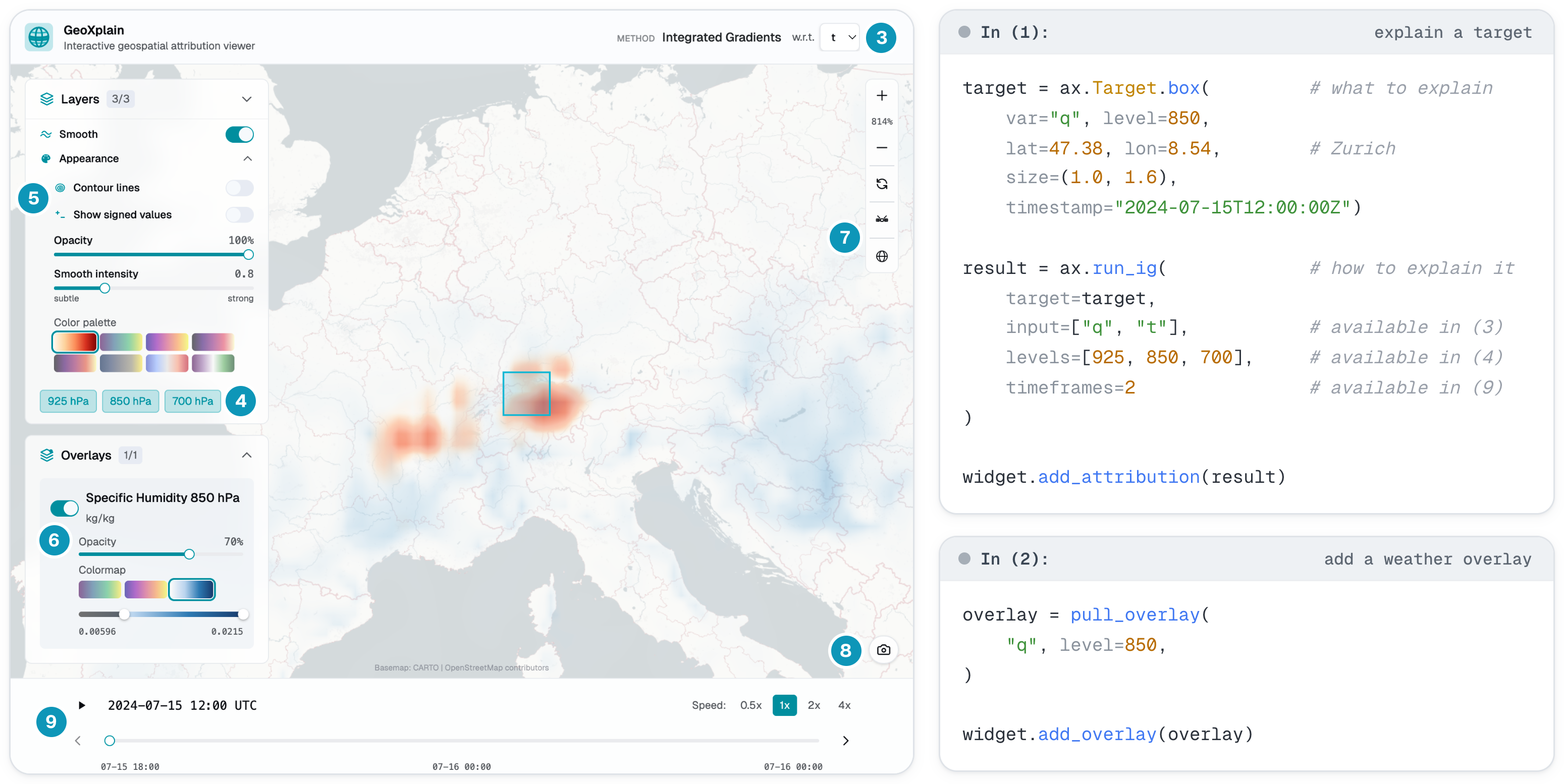}
  \caption{Example of GeoXplain notebook for explaining a humidity forecast at 850\,hPa over Zurich, Switzerland with Integrated Gradients. Attribution data is computed on demand in notebook cell~(1), and a humidity forecast overlay is added in cell~(2). The left panel shows the resulting interactive view. Users can switch between attribution fields for temperature ($t$) and specific humidity ($q$), using control~(3). Individual pressure-level explanations can be shown or hidden with control~(4), and their visual appearance can be adjusted with the panel~(5), for example by switching from filled overlays to contour lines. Panel~(6) adjusts the imported overlay, including its color map and opacity. Controls~(7) switch between topographic and satellite basemaps and enable a 3D globe view. Control~(8) exports the current view as an image, while control~(9) provides timeline controls for exploring explanations across forecast times.}
  \label{fig:teaser}
}

\begin{document}

\firstsection{Introduction}
\maketitle

AI-based weather and earth-system models are rapidly becoming practical scientific instruments \cite{BenBouallegue2024RiseDataDriven}.
These models are attractive for weather- and climate-related workflows because they can produce global forecasts at comparatively low computational cost while remaining competitive with established numerical systems \cite{Bi2023Pangu,Lam2023GraphCast,Bodnar2025FoundationEarthSystem}.
At the same time, their predictions are distributed over space, atmospheric pressure levels, variables, and lead times, which makes it hard to answer basic interpretability questions: Which input fields influence a humid-air forecast over the Alps? Does the answer change across pressure levels? Does it remain stable over an autoregressive rollout? How should an atmospheric scientist compare the attribution with the underlying meteorological field, such as humidity or temperature?

In practice, scientific model-inspection workflows often combine notebook-based exploration, batch computation, and static plotting, producing intermediate outputs that must be inspected, saved, or reused across stages~\cite{Beg2021Jupyter,Hunter2007Matplotlib}.
This separation slows exploratory explanation work~\cite{Yu2024PyGWalker}, especially for climate and weather data, where the analysis must coordinate high-dimensional spatiotemporal fields across views~\cite{Nayeem2024ClimateVisualAnalytics,Diehl2015SpatioTemporalWeather}.
Attribution maps are dense global grids, often sparse, and usually meaningful only in relation to a target region, a forecast timestamp, and contextual meteorological fields.
Analysts therefore need a tool that behaves like exploratory visual analysis software, but remains close to the model code and the high-performance computing environment where explanations are produced.

We introduce \emph{GeoXplain}, an interactive toolkit for visual analysis of weather foundation model explanations.
GeoXplain is designed around a simple separation: the visualization layer displays attribution results, while model-specific adapters compute them.
This paper presents GeoXplain as the main toolkit and the \emph{GeoXplain Aurora Adapter} as a backend that enables on-the-fly computation for Aurora, a leading earth-system foundation model for weather, with applications to air-quality, ocean-wave, and tropical-cyclone prediction~\cite{Bodnar2025FoundationEarthSystem}.

In particular, our contributions are:
(1)~a visualization workflow for dense geospatial attributions over pressure levels and time;
(2)~a self-describing result bundle format that decouples the viewer from the computation package;
(3)~an adaptation of Saliency, Integrated Gradients, RISE, and ViT-CX to Aurora;
(4)~the Aurora adapter that computes the four explanation methods on demand; and
(5)~verification evidence for the rendering and numerical behavior of the current implementation.
Details on the result bundle format, method runners, and diagnostics are provided in Appendices~\ref{appendix:method-runners}--\ref{appendix:aurora-data-flow}.

\section{Related Work}
\label{sec:related-work}

\textbf{Weather foundation models.}
Machine-learning models now match and, on many standard metrics, surpass traditional numerical weather prediction: deterministic systems such as Pangu-Weather, GraphCast, and FourCastNet \cite{Bi2023Pangu, Lam2023GraphCast, Pathak2022FourCastNet}, and probabilistic ones such as GenCast \cite{Price2025GenCast}, operate on atmospheric grids and are evaluated with benchmark suites such as WeatherBench 2~\cite{Rasp2024WeatherBench2}.
This progress has motivated a shift from task-specific models toward foundation models pretrained across heterogeneous variables and tasks~\cite{Nguyen2023ClimaX, Ozdemir2026esfm}. Microsoft Aurora exemplifies this direction as a flexible foundation model for heterogeneous atmospheric data~\cite{Bodnar2025FoundationEarthSystem}.

\textbf{Attribution methods.}
Attribution-based explanation methods help analysts understand model behavior by identifying which input features most influenced a given model output. Gradient-based approaches such as Saliency~\cite{Simonyan2014DeepInside} compute the gradient of a scalar target with respect to the input, while Integrated Gradients~\cite{Sundararajan2017Axiomatic} integrates such gradients along a path from a baseline input to the observed input.
Perturbation-based approaches such as RISE~\cite{Petsiuk2018RISE} estimate feature influence by randomly masking parts of the input and measuring the resulting changes in model output.
Transformer-oriented methods adapt attribution to tokenized internal representations. ViT-CX~\cite{Xie2023VitCX} explains vision-transformer predictions through clustered spatial features and their effects on the output.
GeoXplain itself is agnostic to the type of explanation used: each model adapter defines how an attribution is computed, and the viewer turns the resulting fields into comparable geospatial visual objects. The adapter introduced in this paper provides Saliency, Integrated Gradients, RISE, and ViT-CX for Aurora.

Such methods have been used for weather-model forecast sensitivity, precipitation and severe-weather prediction, and broader geoscientific model analysis \cite{Espeholt2022MetNet2,BanoMedina2025Xynthia,Hua2026PanguSevere,Mamalakis2022Fidelity,Bommer2024FindingXAI}.
Recent work has also begun to analyze Aurora with representation analysis and layer-wise relevance propagation~\cite{Kasteleyn2026AuroraLRP}.
To the best of our knowledge, GeoXplain is the first publicly available system to adapt Saliency, Integrated Gradients, RISE, and ViT-CX to Aurora.
This required several Aurora-specific modifications of the methods, which are documented in Appendix~\ref{appendix:method-runners}.
Since visual plausibility alone is not a reliable validation criterion for attribution maps, our diagnostics include sanity-check practices such as model-parameter randomization~\cite{Adebayo2018SanityChecks}.

\textbf{Weather, climate, and ocean visualization.}
Visualization systems for geophysical forecasting emphasize domain-specific coordinates, time, vertical structure, and uncertainty.
Met.3D provides interactive 3D analysis of numerical ensemble weather forecasts, combining meteorological 2D conventions with 3D views and ensemble support \cite{Rautenhaus2015Met3D}.
Ovis similarly supports interactive analysis of ocean forecast ensembles with derived quantities recomputed as users explore space, time, and ensemble subsets \cite{Hollt2014Ovis}.
GeoXplain shares the concern for preserving geophysical structure, but its primary data are attribution fields for learned models rather than forecast fields or ensemble members.

\textbf{Interactive ML-explanation visualization.}
VIS systems have long treated model interpretation as an interactive visual analysis problem, as surveyed by Hohman et al.~\cite{Hohman2019VisualAnalytics}.
ActiVis uses coordinated views for instance- and subset-level exploration of large neural networks \cite{Kahng2018ActiVis}, while Summit summarizes activations and attributions into scalable interactive representations of learned features and their influence \cite{Hohman2020Summit}.
GeoXplain follows this line of work, but shifts the visual object to georeferenced attribution fields. Its contribution is a domain-specific visualization design and result protocol for comparing explanations across variables, pressure levels, forecast times, and physical overlays.

\textbf{Notebook-embedded visual analysis.}
Computational-notebook systems such as B2 \cite{Wu2020B2} and PyGWalker \cite{Yu2024PyGWalker} reduce the gap between code-driven analysis and interactive visual exploration by making visual interfaces callable from the analysis environment.
GeoXplain follows this on-the-fly pattern; however, attribution computations for foundation models are expensive and may have to be computed on a GPU or SLURM job.
This motivates our explicit split between a reusable visualization protocol and model-specific computation adapters, enabling full flexibility in what to compute while maintaining smooth, uninterrupted interactions with the visualizations.

\section{GeoXplain}

\subsection{Overview}

\begin{figure*}[t]
  \centering
  \includegraphics[
    width=0.9\textwidth,
    height=3.7in,
    keepaspectratio
  ]{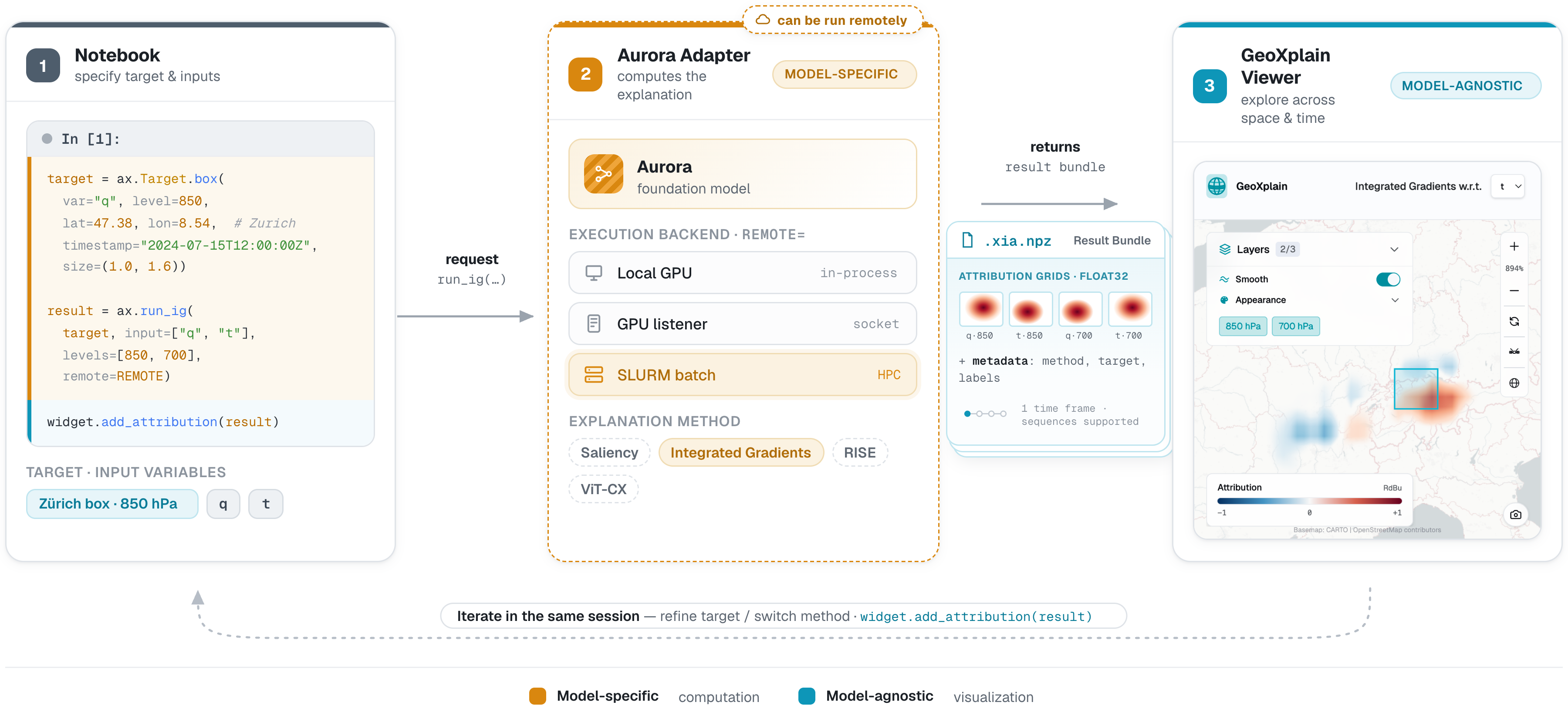}
  \caption{GeoXplain separates interactive visual analysis from model-specific computation. In the notebook (1), the analyst specifies the target and input variables.
The model-specific Aurora adapter (2) turns this target specification into a result bundle containing attribution time frames and metadata.
The model-agnostic GeoXplain viewer (3) renders the bundle in a notebook widget or standalone browser view.
This example shows how the specific humidity at 850\,hPa around Zurich is explained using humidity and temperature at 850 and 700\,hPa.}
  \label{fig:overview}
\end{figure*}

GeoXplain provides two user interfaces: a standalone browser viewer and a Jupyter notebook widget that directly integrates the viewer into the development workflow using~\texttt{anywidget}~\cite{Manz2024}.
Both interfaces share the same data model, preprocessing, and encoding pipeline, illustrated in Fig.~\ref{fig:overview}.
The frontend provides map and globe views, method selection, timeline playback, pressure-level visibility and opacity controls, heatmap or contour rendering, palette selection, target annotations, and optional overlays for physical variables such as specific humidity or temperature (see Fig.~\ref{fig:teaser}). Implementation details can be found in Appendix~\ref{appendix:geoexplain-implementation}.

\subsection{Design Goals}

The design follows four analytics goals.
\emph{G1: Uninterrupted development workflow.}
Users should be able to inspect their models without interrupting their development workflow. Many researchers use computational notebooks for their experiments. Integrating the visual analysis within the notebook environment maintains development flow and avoids the common detour through job logs, copied arrays, and static plotting scripts.
\emph{G2: Preserve weather-specific context.}
Attribution values should remain tied to their meteorological meaning, including variable, forecast time, target location, and vertical pressure level, rather than being shown as unlabeled array dimensions.
\emph{G3: Compare explanation and context.}
Attribution maps and raw physical-field overlays should be shown in the same coordinate system since meteorological interpretation depends on both.
\emph{G4: Isolate model-specific code.}
The visualization package should be agnostic to the specific model and attribution method to provide a lightweight, reusable system.

These goals shaped GeoXplain's overall design, including its model-agnostic result bundle format.
A GeoXplain result bundle groups attribution grids with the metadata needed to render them, including the explanation method, time frame, target, input variables, and vertical layers.
The same format is used by the notebook widget, standalone browser viewer, and remote computation backends. Storage details are provided in Appendix~\ref{appendix:result-format}.

\subsection{Usage Scenario}
In GeoXplain, a target defines the forecast output to be explained. It includes the output variable, forecast time, optional vertical level, and spatial region. A point target explains the output at a single coordinate location, whereas a box target first averages the output over a latitude--longitude rectangle and then computes the attribution.

Consider an atmospheric scientist investigating a warm, humid summer case over Zurich, Switzerland, explaining Aurora's subsequent 850\,hPa specific-humidity forecast from the 2024-07-15~12:00~UTC input state.
In a notebook, the scientist specifies a box target for specific humidity at 850\,hPa ($\approx1500\,\mathrm{m}$ above sea level) and selects input variables such as temperature, humidity, and geopotential.
The call
\texttt{run\_ig(target, input=[...], remote=REMOTE)}
submits an Integrated Gradients computation to the specified backend.
When the result returns, \texttt{w.add\_attribution(result)} places each input variable in the method selector and each returned frame on the timeline.
The scientist can then switch from the humidity attribution to temperature attribution, compare 850\,hPa with higher atmospheric layers, adjust opacity and view contour lines.

\begin{figure*}[t]
  \centering
  \includegraphics[
    width=0.94\textwidth,
    height=3.7in,
    keepaspectratio
  ]{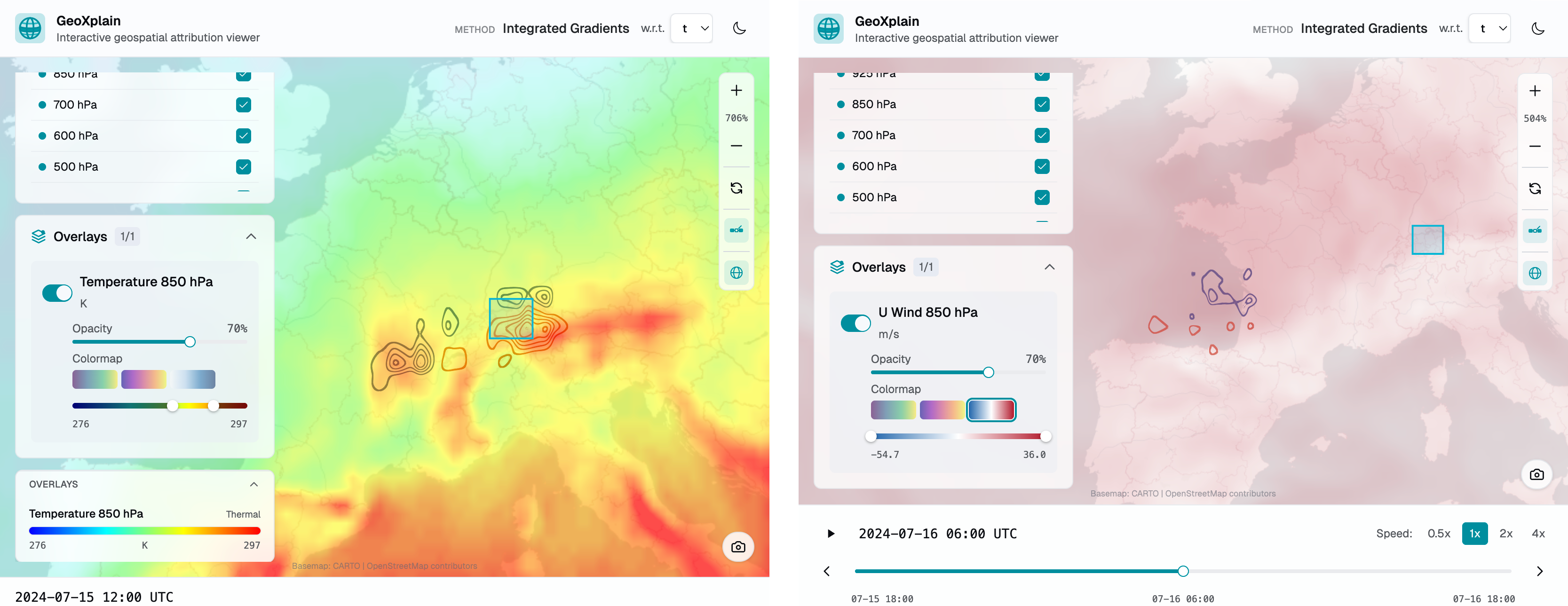}
  \caption{The left image shows IG attributions for the 850\,hPa humidity forecast over Zurich with an 850\,hPa temperature overlay. The right image shows the second autoregressive rollout step with the 850\,hPa horizontal wind field. Positive wind values indicate an eastward flow. In both panels, attribution is shown by contour lines, with red indicating positive and blue indicating negative attribution.}
  \label{fig:example}
\end{figure*}

The initial IG result already provides a first hypothesis. As shown in the left panel of Fig.~\ref{fig:example}, the attribution map for the Zurich humidity target assigns positive contribution to 850\,hPa temperature southeast of Zurich and negative contribution to a region near Lyon. This suggests that Aurora's humidity prediction is sensitive not only to local humidity, but also to nearby thermal structure and upstream conditions.

The same session can then be extended. To test whether this pattern is method-specific, the scientist requests RISE for the same target. RISE highlights similar regions, which increases confidence that the observed patches are not merely an artifact of the IG baseline path.

To inspect the meteorological context, the scientist adds physical overlays. With \texttt{pull\_overlay("q", level=850, remote=REMOTE)}, the 850\,hPa humidity field is retrieved from the remote backend. Metadata such as the default color map are inferred from the request, and the adapter provides the previously used timestamps automatically. After calling \texttt{w.add\_overlay(overlay)}, the scientist adds the field to the running viewer and repeats the same step for temperature. The attributed regions, especially the one to the south, coincide with slightly warmer 850\,hPa air, suggesting that the model may use local thermal structure as part of the humidity forecast. However, this visual agreement alone is not sufficient evidence for a physical mechanism. 

To see whether the same regions remain influential over later forecast steps, the scientist requests an autoregressive rollout with \texttt{run\_rollout(..., method="ig", timeframes=5)}, which iteratively uses each forecast state as input for the subsequent forecast step and computes each IG attribution with respect to the original input state. GeoXplain displays the returned attribution frames as a playable sequence. Across longer lead times, the initial patches weaken, while stronger attribution appears farther upstream over western France and northern Spain, as shown in the right panel of Fig.~\ref{fig:example}.

The scientist then adds eastward and northward wind-component overlays, using custom color maps with \texttt{w.add\_overlay(overlay, colormap=...)} to separate them visually. The wind fields show persistent northeastward flow from northern Spain and western France toward the Zurich region. This makes the attribution pattern meteorologically plausible: the model may be responding to upstream air-mass transport rather than only to local humidity or temperature anomalies.

The scientist then uses the export-as-image feature to export the current view for documentation and calls \texttt{result.save(...)} to store the explanations as a result bundle that can be shared with colleagues.

The notebook remains the provenance record for targets and method settings, while the viewer provides the interactive surface for spatial comparison. This workflow follows the notebook-embedded visual analysis pattern discussed in Section~\ref{sec:related-work} and illustrated in Fig.~\ref{fig:overview}: one line activates the tool, while the analyst continues refining targets and method settings in code.

\subsection{Visual Representation}

GeoXplain represents attributions as dense global grids on a latitude--longitude domain, rendered as filled heatmaps or contour lines on a geospatial map. Heatmaps emphasize attribution magnitude, while contour lines can be overlaid with other layers without competing color scales.
Signed variables are automatically detected as diverging and displayed with diverging palettes centered at zero by default. Nonnegative variables default to sequential palettes.
Users can switch between heatmaps and contour-line depictions and can seamlessly change between a map view and 3D globe view, as shown in Fig.~\ref{fig:teaser}.

Physical-field overlays are stored separately from attribution layers.
An overlay has its own timestamps, units, and color map, allowing users to compare an explanation map with the physical field that was explained or with another contextual variable.
Because overlays use the same timeline as explanations, they can accompany multi-step rollouts and batch time frames.

\section{On-the-fly Computation}
GeoXplain uses adapters as computation backends between model code and the viewer. An adapter translates a target and explanation request into GeoXplain's standardized result bundle, keeping the visualization package independent of any particular model implementation, data loader, attribution method, or execution environment. In practice, adapters resolve targets, prepare model inputs, run the requested explanation or overlay computation, and package attribution grids and metadata in the common format.

\subsection{Aurora Adapter}

The Aurora adapter implements this backend pattern for Aurora and WeatherBench-style case data.
A target defines the output variable, pressure level, timestamp, and either a point or a latitude-longitude box (Fig.~\ref{fig:overview},~(1)).
The adapter loads the matching WeatherBench-style case data, builds an Aurora batch, constructs a differentiable scalar target, runs the selected explanation method, and returns attribution maps for the requested input variables, as shown in Fig.~\ref{fig:overview},~(2).
For a single timestamp, the result contains one frame.
For batch time frames or rollouts, it contains multiple frames that GeoXplain places on the timeline.

The same Python call can be executed locally, through a GPU listener, or through a SLURM-backed listener, corresponding to the execution backends in Fig.~\ref{fig:overview},~(2).
Remote requests use status polling with progress and log tails, and return a result bundle when complete.
This addresses a practical mismatch between interactive visual analysis and HPC (high-performance computing) execution: the analyst wants to make small target or method changes in a notebook, while the model may need a GPU node and containerized runtime.
The adapter hides these constraints behind the \texttt{remote} argument, and the returned result object is identical in all execution modes.

\subsection{Data Flow}

The adapter computes the results for each attribution method and returns the outputs in the standardized result bundle format.
For each request, the adapter prepares the Aurora batch, resolves the target to a scalar output, and masks the requested fields for attribution.
It then converts the returned arrays into GeoXplain frames including the variable, pressure level, color map, and runtime metadata.
Batch time frames repeat this process over consecutive timestamps and merge the frames into one bundle.
Rollout mode instead backpropagates through an autoregressive sequence, so a single target can ask which initial inputs mattered for later lead times. Further implementation details are provided in Appendix~\ref{appendix:aurora-data-flow}.

\subsection{Adapting Other Models}

GeoXplain's model-specific interface is intentionally narrow.
A new model backend must provide (1) a target specification that resolves to a scalar model output, (2) a case loader and batch builder, (3) method runners that produce layer grids, and (4) a serializer that satisfies the result bundle format.
Nothing in the GeoXplain viewer requires Aurora-specific imports.
This separation is important because the same visualization questions arise for regional models, climate emulators, downscaling models, and foundation models.
For example, regional precipitation or climate-emulator adapters could reuse the viewer while replacing only the target resolver, data loader, model wrapper, and layer labels.
The visual comparison tasks remain the same: select a target, inspect the attribution field, compare across variables and time, and check the explanation against physically meaningful overlays.

\section{Verification}

The viewer is tested with visual parity checks that render synthetic ramps, blobs, and checkerboards, capture the output, and compare screenshots against Matplotlib references for color maps, opacity, and smoothing.
The adapter has unit tests for API routing, batch/time-frame expansion, remote polling, overlay serialization, and login-node overlay execution.
Numerical diagnostics were run on four GH200 GPUs with the public Aurora 6 h model, using multiple target cases for randomization and method-specific checks.
Integrated Gradients showed sub-percent to about 1\% completeness residuals in most cases, and full-model randomization removed learned attribution structure.
We treat these results as necessary sanity checks rather than a complete diagnostic analysis. Details are provided in Appendices~\ref{appendix:xai-methods-verification} and~\ref{appendix:viewer-verification}.

\section{Conclusion and Future Work}

GeoXplain connects interactive geospatial visualization with on-the-fly explanation computation for weather and climate foundation models.
Its Aurora adapter demonstrates the proposed result protocol, which can be adapted to other models and workloads.
The central contribution is an accessible way to inspect, compare, and discuss model explanations in their geophysical context.

We plan to evaluate the visualization design with weather or climate scientists in future work.
Costly ViT-CX computations and long rollouts can cause substantial waiting times. GeoXplain separates computation from interaction to reduce workflow disruption, but studies are needed to test whether this improves the analysis experience or whether alternative strategies are required.
Finally, the current protocol emphasizes dense regular grids optimized for Aurora and WeatherBench-style data.
Supporting unstructured meshes, station observations, ensembles, uncertainty fields, and multi-resolution regional nests will require protocol extensions.

\section*{Software Availability}
\label{sec:software_availability}

GeoXplain is released as two open-source Python packages, \texttt{geoxplain} and \texttt{geoxplain-aurora-adapter}, installable via PyPI. Source code, documentation, and runnable examples are available at the project repository \url{https://github.com/clemenskoprolin/geoxplain}.

\bibliographystyle{abbrv-doi-hyperref}
\bibliography{geoxplain}

\clearpage
\appendix

\section{Method Runner Adaptations}
\label{appendix:method-runners}

All method runners explain the same scalar target. Let \(X\) denote the Aurora input case and let
\[
S(X) = \tau(F(X))
\]
be the scalar obtained by applying Aurora \(F\) and resolving the user target \(\tau\): either a point value or an arithmetic grid-cell mean over a target box, at a selected pressure level for atmospheric variables.

Requested input variables are denoted by \(v\in\mathcal{V}\). Atmospheric fields include two input time steps, pressure levels, and the horizontal grid, while surface fields omit the pressure-level dimension. We write \(t_1\) for the second input time step, \(X'\) for a smoothed-baseline case, and \(\delta=10^{-8}\) for guarded divisions.

The exported attribution is always with respect to the second input time step, \(t_1\), because the requested timestamp is interpreted as Aurora's current input state. Preliminary testing showed that this choice produced more interpretable attribution maps, although future work could also consider attributions with respect to the earlier input state \(t_0\).

\textbf{Saliency.}
For each requested variable, saliency is the gradient of the scalar target with respect to the selected input field,
\[
A_v^{\mathrm{sal}}
=
\frac{\partial S(X)}{\partial X_{v,t_1}} .
\]
All requested variables are marked as differentiable inputs, and their gradients are computed in a single backward pass. Atmospheric gradients are then split into pressure-level maps, while surface gradients are stored under the reserved \texttt{sfc} layer.

\textbf{Integrated Gradients.}
Integrated Gradients attributes the target by accumulating gradients along a path from a smoothed baseline \(X'\) to the actual input \(X\). In this adapter, \(X'\) is obtained by Gaussian smoothing the input fields in latitude and longitude, using reflected latitude boundaries and wrapped longitude boundaries.

For each requested variable,
\[
A_v^{\mathrm{IG}}
=
(X_{v,t_1}-X'_{v,t_1})
\frac{1}{n}
\sum_{k=1}^{n}
\frac{\partial S(X^{(k)})}{\partial X^{(k)}_{v,t_1}},
\qquad
X^{(k)} = X' + \frac{k-0.5}{n}(X-X') ,
\]
where \(n\) is the number of integration steps. The points \(X^{(k)}\) are midpoint samples on the straight path from \(X'\) to \(X\).

For selected variables, the IG path interpolates the selected variable fields, including both input time steps, and the \(t_1\) attribution is exported. Unselected fields remain fixed at their original input values. The target resolver and export convention are the same as for saliency. Future work could investigate alternative baselines or other integration paths.

\textbf{RISE.}
RISE estimates attribution by perturbing one requested input variable at a time and measuring how the scalar target changes. In this adapter, perturbations are applied only to the second input time step \(t_1\), while \(t_0\) is kept fixed.

For a random horizontal keep mask \(M^{(k)}\), the selected field is blended with its smoothed baseline,
\[
\tilde X_{v,t_1}^{(k)}
=
M^{(k)} \odot X_{v,t_1}
+
(1-M^{(k)}) \odot X'_{v,t_1},
\]
where \(\odot\) denotes element-wise multiplication. The modified case \(\tilde X_v^{(k)}\) is otherwise identical to \(X\). For atmospheric variables, the same horizontal mask is applied at all pressure levels, so the perturbation is column-wise. The masks are generated from low-resolution Bernoulli grids, then bilinearly upsampled into continuous masks in \([0,1]\) and shifted circularly in longitude to avoid dateline artifacts.

Rather than using the original classification-style RISE normalization, the adapter uses a centered regression estimator,
\[
A_{v,i}^{\mathrm{RISE}}
=
\begin{cases}
\displaystyle
\frac{
\sum_{k=1}^{N} (S(\tilde X_v^{(k)})-\bar S)(M_i^{(k)}-\bar M_i)
}{
\sum_{k=1}^{N} (M_i^{(k)}-\bar M_i)^2
},
&
\mathrm{var}(M_i)>\delta,\\[1.2em]
0, & \mathrm{otherwise},
\end{cases}
\]
where \(i\) indexes a horizontal grid cell, \(N\) is the number of masks,
\[
\bar S = \frac{1}{N}\sum_{k=1}^{N} S(\tilde X_v^{(k)}),
\qquad
\bar M_i = \frac{1}{N}\sum_{k=1}^{N} M_i^{(k)} .
\]
This estimator removes the constant forecast offset and assigns positive values to regions whose visibility is associated with a higher scalar target. Because atmospheric masks are applied column-wise, atmospheric variables are exported as one collapsed horizontal attribution map per variable.

\textbf{ViT-CX.}
ViT-CX requires an Aurora-specific adaptation because Aurora's encoder is hierarchical and does not expose a single fixed image-patch grid. The runner therefore hooks a user-selected encoder stage, \texttt{hook\_stage}. Each stage has its own horizontal token resolution, feature dimension, and compressed vertical token groups.

For stage 2, the recorded activation is reshaped as
\[
Z \in \mathbb{R}^{G \times H_e \times W_e \times D},
\qquad
\bar Z_{ij:} = \frac{1}{G}\sum_{g=1}^{G} Z_{gij:},
\]
where \(G=4\) denotes the vertical token groups, \((H_e,W_e)=(45,90)\) is the encoder-grid resolution, and \(D\) is the feature dimension. The vertical groups are averaged so that each horizontal encoder cell is represented by one feature vector \(\bar Z_{ij:}\).

Perturbation regions are then obtained by clustering the normalized spatial token vectors \(\bar Z_{ij:}\). This differs from the original ViT-CX formulation, which constructs masks from embedding-channel feature maps. The clustering produces masks \(C_1,\ldots,C_K\) on the encoder grid, with \(K\) set by \texttt{n\_clusters}. Each mask represents a spatial region. For atmospheric variables, the corresponding horizontal region is applied to all pressure levels.

Each cluster mask is upsampled to the full input grid and used as an occlusion region. Let \(X_{v}^{(k,\mathrm{occ})}\) denote the modified case in which the selected \(t_1\) field of variable \(v\) is replaced by its smoothed baseline inside cluster \(C_k\). The cluster score is
\[
s_k = S(X) - S\!\left(X_{v}^{(k,\mathrm{occ})}\right).
\]
Positive scores therefore indicate regions whose removal decreases the target.

The final attribution map is formed by assigning each cluster score back to its spatial support,
\[
A_v^{\mathrm{CX}}(i)
=
\begin{cases}
\displaystyle
\frac{\sum_{k=1}^{K} s_k\,\mathbf{1}\{i\in C_k\}}
     {\sum_{k=1}^{K} \mathbf{1}\{i\in C_k\}},
&
\sum_{k=1}^{K} \mathbf{1}\{i\in C_k\}>\delta,\\[1.2em]
0, & \mathrm{otherwise},
\end{cases}
\]
where \(i\) indexes a horizontal grid cell. The resulting encoder-grid map is upsampled to the full grid and, optionally, smoothed. As in RISE, atmospheric variables are perturbed column-wise and therefore produce one collapsed horizontal attribution map per input variable.

\textbf{Rollout.}
Rollout changes the forecast horizon while keeping the attribution input fixed. For rollout frame \(r\), Aurora is applied autoregressively for \(r+1\) six-hour steps and the scalar target is evaluated on the final prediction,
\[
S_r(X) = \tau(F^{\,r+1}(X)).
\]
Saliency or Integrated Gradients is then computed for \(S_r\) with respect to the original input case \(X\), not with respect to the intermediate predicted states. Thus each rollout frame asks how the initial input state contributes to the selected target at a later lead time.

The returned result bundle stores one frame per lead time, with metadata for rollout index, lead hours, runtime, and target score. RISE and ViT-CX are excluded from rollout because their perturbation loops would require many full autoregressive forwards per variable. Future work could investigate this.

\section{XAI Methods Verification}
\label{appendix:xai-methods-verification}

\textbf{Diagnostic setup.}
The following checks are intended as implementation diagnostics, not as an evaluation of forecast skill or as a comparison against meteorological baselines. They used the public Aurora 6\,h model and were run across four GH200 GPUs. The shared diagnostic parameters are summarized in Table~\ref{tab:xai-diagnostic-parameters}. The randomization checks used three target cases: an Alpine land \(q@850\) box, a North Atlantic \(t@500\) box, and a tropical Pacific \(q@700\) box. The broader Integrated Gradients completeness suite also included additional seasonal and variable-set variants. Table~\ref{tab:xai-diagnostic-cases} lists the target geometry and attribution variables for all reported cases. These checks are sanity tests: passing them is necessary for trustworthy attribution, but does not by itself establish scientific validity. The diagnostic scripts are available for reproducibility at \url{https://github.com/clemenskoprolin/geoxplain-aurora-adapter/tree/master/diagnostics}.

\begin{table}[H]
\centering
\small
\setlength{\tabcolsep}{3pt}
\caption{Main parameters of the numerical diagnostics.}
\label{tab:xai-diagnostic-parameters}
\begin{tabular}{@{}>{\raggedright\arraybackslash}p{0.27\columnwidth}>{\raggedright\arraybackslash}p{0.68\columnwidth}@{}}
\toprule
Item & Value \\
\midrule
Model & Public Aurora 6\,h pretrained checkpoint (\texttt{microsoft/aurora}) \\
  IG baseline & Gaussian-smoothed input, \(\sigma=2.5^\circ\) \\
  IG step counts & \(4,8,16,32,64\) midpoint samples \\
  RISE parameters & An \(18\times36\) low-resolution mask grid \\
  Randomization seeds & \(11,23,37,51,67\) for full randomization \\
  Cascade seeds & \(11,23\), output-to-input cumulative randomization \\
  RISE convergence seeds & Mask seeds: clean A \(=42\), clean B \(=1234\) \\
  RISE convergence checkpoints & \(32,64,128,256,512,1024\) masks, same \(18\times36\) grid \\
  ViT-CX parameters & Encoder stage \(2\), with token grid \(4\times45\times90\) \\
\bottomrule
\end{tabular}
\end{table}

\begin{table*}[tbp]
\centering
\scriptsize
\setlength{\tabcolsep}{4pt}
\caption{Diagnostic cases and target geometry. Longitudes use the ERA5 \(0^\circ\)--\(360^\circ\) convention. Box sizes are latitude by longitude in degrees.}
\label{tab:xai-diagnostic-cases}
\begin{tabular}{@{}lllllll@{}}
\toprule
Case & Init & Target & Mode & Center/point \((\phi,\lambda)\) & Size & Attrib. vars \\
\midrule
\texttt{alps\_w} & 2024-01-15 00Z & \(q@850\) & box & \((46.25, 8.75)\) & \((1.5,2.5)\) & \(t,q,z\) \\
\texttt{alps\_p} & 2024-04-10 12Z & \(q@850\) & box & \((46.25, 8.75)\) & \((1.5,2.5)\) & \(t,q,z\) \\
\texttt{alps\_s} & 2024-07-15 00Z & \(q@850\) & box & \((46.25, 8.75)\) & \((1.5,2.5)\) & \(t,q,z\) \\
\texttt{alps\_a} & 2024-10-15 12Z & \(q@850\) & box & \((46.25, 8.75)\) & \((1.5,2.5)\) & \(t,q,z\) \\
\texttt{alps\_w\_q} & 2024-01-15 00Z & \(q@850\) & box & \((46.25, 8.75)\) & \((1.5,2.5)\) & \(q\) \\
\texttt{alps\_w\_all} & 2024-01-15 00Z & \(q@850\) & box & \((46.25, 8.75)\) & \((1.5,2.5)\) & \(t,q,z,u,v\) \\
\texttt{atl\_s} & 2024-07-15 00Z & \(t@500\) & box & \((50.00, 330.00)\) & \((3.0,4.0)\) & \(t,q,z\) \\
\texttt{trop\_s} & 2024-07-15 00Z & \(q@700\) & box & \((5.00, 200.00)\) & \((4.0,5.0)\) & \(q\) \\
\texttt{pt\_s} & 2024-07-15 00Z & \(z@500\) & point & \((40.00, 280.00)\) & -- & \(t,q,z,u,v\) \\
\bottomrule
\end{tabular}
\end{table*}

\textbf{Integrated Gradients completeness.}
For Integrated Gradients, the main numerical diagnostic was completeness. The baseline \(X'\) was constructed by Gaussian smoothing the input fields in latitude and longitude, and the path integral was approximated with midpoint samples. We compared the summed attribution with the target-score difference \(S(X)-S(X')\). The integration path interpolates both input time steps, so the sum runs over \(t_0\) and \(t_1\) jointly. The per-variable \(t_1\) slice used
for visualization is only a part of this total, and is not expected to satisfy completeness on its own. Table~\ref{tab:xai-ig-completeness} reports the 64-step results for the full completeness suite: seasonal Alpine \(q@850\) boxes, single-, three-, and five-variable attribution sets, and three auxiliary non-\(q@850\) targets. For the Alpine \(q@850\) box targets, residuals were sub-percent except for the winter \(q\)-only run, which reached \(1.02\%\). The auxiliary cases ranged from \(0.02\%\) for tropical \(q@700\) to \(2.97\%\) for the point \(z@500\) target, where the absolute gap was still small relative to the attribution energy. Thus, for the \(q@850\) box setting, IG completeness can be summarized as sub-percent to about \(1\%\), with the wider diagnostic suite reaching about \(3\%\) in the point-target case.

\begin{table}[H]
\centering
\scriptsize
\setlength{\tabcolsep}{1.5pt}
\caption{Integrated Gradients completeness at 64 midpoint samples. Gap and energy are relative percentages.}
\label{tab:xai-ig-completeness}
\begin{tabular}{@{}llrrrr@{}}
\toprule
Case & Target & \(\Delta\) & \(\Sigma\)IG & gap & energy \\
\midrule
\texttt{alps\_w} & \(q@850\) & \(-2.71{\times}10^{-4}\) & \(-2.72{\times}10^{-4}\) & 0.31 & 0.018 \\
\texttt{alps\_p} & \(q@850\) & \(-1.25{\times}10^{-4}\) & \(-1.26{\times}10^{-4}\) & 0.58 & 0.009 \\
\texttt{alps\_s} & \(q@850\) & \(4.30{\times}10^{-4}\) & \(4.30{\times}10^{-4}\) & 0.05 & 0.002 \\
\texttt{alps\_a} & \(q@850\) & \(2.48{\times}10^{-4}\) & \(2.47{\times}10^{-4}\) & 0.49 & 0.014 \\
\texttt{alps\_w\_q} & \(q@850\) & \(-2.76{\times}10^{-4}\) & \(-2.79{\times}10^{-4}\) & 1.02 & 0.181 \\
\texttt{alps\_w\_all} & \(q@850\) & \(-3.14{\times}10^{-4}\) & \(-3.15{\times}10^{-4}\) & 0.35 & 0.015 \\
\texttt{atl\_s} & \(t@500\) & \(5.19{\times}10^{-1}\) & \(5.21{\times}10^{-1}\) & 0.39 & 0.013 \\
\texttt{trop\_s} & \(q@700\) & \(-1.05{\times}10^{-3}\) & \(-1.05{\times}10^{-3}\) & 0.02 & 0.005 \\
\texttt{pt\_s} & \(z@500\) & \(7.94\) & \(7.71\) & 2.97 & 0.003 \\
\bottomrule
\end{tabular}
\end{table}

\textbf{Parameter randomization.}
The parameter-randomization sanity check followed Adebayo et al.: attributions from the trained model were compared with attributions after progressively randomizing model components. First, cascading randomization progressively randomized parameter groups from the Perceiver decoder back toward the input encoder. Attribution similarity and spatial energy decreased as more of the network was randomized, and all methods reached full collapse once the encoder and input tokenizer were included, with energy ratios approaching zero and 100\% collapse at the final stage. Second, Table~\ref{tab:xai-full-randomization} reports the full-model randomization endpoint over the three target cases and five random seeds. IG, RISE, and ViT-CX collapsed to near-constant attribution fields, while saliency retained only negligible spatial energy. The mean energy ratio was about \(6\times10^{-5}\) for IG and RISE, \(2\times10^{-4}\) for ViT-CX, and \(7\times10^{-4}\) for saliency. Correlation-based similarities were therefore at or below the calibrated spatial-noise baseline. This supports the sanity requirement that the maps depend on the learned Aurora parameters rather than only on fixed preprocessing, target geometry, or visualization choices.

\begin{table}[H]
\centering
\scriptsize
\setlength{\tabcolsep}{3pt}
\caption{Full-model randomization over three cases and five seeds.}
\label{tab:xai-full-randomization}
\begin{tabular}{@{}lrrcr@{}}
\toprule
Method & Runs & Collapse & Pearson & Energy ratio \\
\midrule
Saliency & 15 & 10/15 & \(0.009 \pm 0.021\) & \((6.95 \pm 7.6){\times}10^{-4}\) \\
IG & 15 & 15/15 & -- & \((5.85 \pm 3.9){\times}10^{-5}\) \\
RISE & 15 & 15/15 & -- & \((6.34 \pm 4.7){\times}10^{-5}\) \\
ViT-CX & 15 & 15/15 & -- & \((1.93 \pm 1.7){\times}10^{-4}\) \\
\bottomrule
\end{tabular}
\end{table}

\textbf{RISE Monte Carlo convergence.}
RISE was additionally checked for Monte Carlo convergence using the mask counts listed in Table~\ref{tab:xai-diagnostic-parameters}. With 32 to 1024 masks, the self-consistency of the estimate increased monotonically on the tested cases, with mean self-correlation rising from about \(0.30\) to \(1.00\). Agreement between independent mask draws also improved, from about \(0.06\) to \(0.68\) mean Pearson correlation. This confirms the expected variance reduction with more perturbation samples, while also showing why RISE is more expensive and noisier than the gradient methods at low mask counts.

\textbf{Summary and scope.}
Taken together, these diagnostics indicate that IG has the expected completeness behavior for the tested targets, RISE shows the expected sample-count dependence, and full-model randomization removes the learned structure from the attribution maps. However, the suite only covers nine case variants, and most methods were tested with only one parameter setting. We therefore treat these results as preliminary implementation evidence and leave broader tests across variables, targets, locations, method settings, and input-perturbation robustness to future work.

\section{Viewer Verification}
\label{appendix:viewer-verification}
\textbf{Verification setup.}
The viewer is checked at two complementary levels. Protocol tests exercise the data handoff from Python into the standalone browser viewer, notebook widget, and remote backends. These tests focus on whether targets, overlays, timelines, input variables and other metadata are serialized consistently. A separate visual parity suite then checks whether the browser rendering of those payloads has the intended appearance.

For the visual checks, synthetic attribution fields are created and prepared for viewing. The suite renders independent Matplotlib references, starts the viewer, and captures the map output. Each viewer capture is generated from a fixed launch state, so the map camera, map style, visible layer, and smoothing options are controlled. The run writes reference images, viewer screenshots, and side-by-side comparison panels for inspection.

\textbf{Test cases.}
Each synthetic test isolates a visible property of the viewer. West-to-east ramps test the diverging and sequential color maps. Gaussian blobs test how weak and strong attribution values fade or saturate under the opacity rule. Alternating positive and negative stripes test the near-zero transparent band in the signed display. Checkerboards and single-cell spikes test imported-grid smoothing with smoothing disabled and enabled.

The Matplotlib reference renderer follows the same normalization and opacity conventions as the viewer, while remaining independent of the browser implementation. Some differences are expected: the reference image uses a neutral background, whereas the viewer is drawn over MapLibre topographic tiles. The reference is shown on a longitude--latitude grid, whereas the viewer uses Web Mercator.

\begin{figure*}[tbp]
  \centering
  \includegraphics[
    width=0.98\textwidth,
    height=4.1in,
    keepaspectratio
  ]{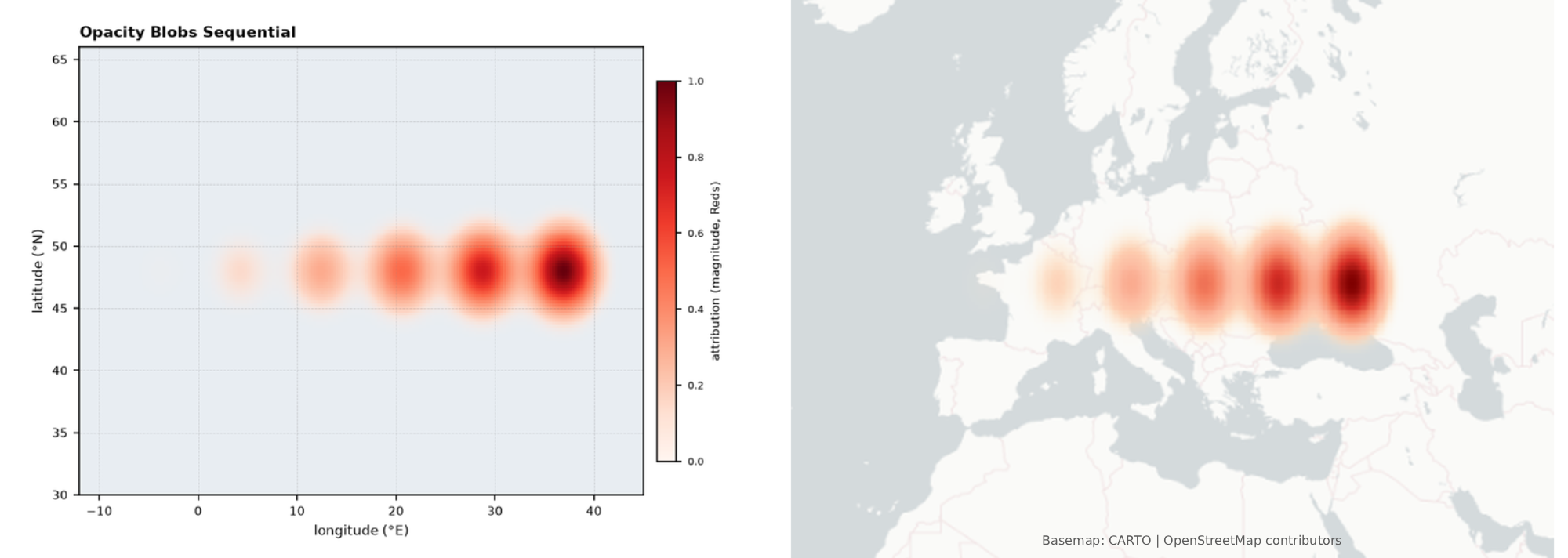}
  \caption{Example output from the viewer visual parity suite. The left panel is the independent Matplotlib reference for the sequential opacity-blob case, and the right panel is a capture of the GeoXplain viewer under the matching launch state. This case checks the sequential color map and opacity transfer: weak blobs remain faint, while stronger blobs saturate toward deep red over the map background.}
  \label{fig:viewer-verification-compare}
\end{figure*}

\textbf{Results.}
The protocol tests and visual parity cases behave as expected. The exported payloads are accepted consistently by the viewer entry points, and the browser captures reproduce the intended color maps, transparency behavior, sign handling, and smoothing choices. The generated comparison panels show agreement with the Matplotlib references apart from the documented background, projection, and cropping differences. Fig.~\ref{fig:viewer-verification-compare} shows one representative comparison for the sequential opacity-blob case. The verification suite is provided at \url{https://github.com/clemenskoprolin/geoxplain/tree/main/tests/visual_test} for reproducibility.

\section{GeoXplain Implementation}
\label{appendix:geoexplain-implementation}
The \texttt{geoxplain} package keeps the visualization layer independent of any computation backend.
\texttt{add\_attribution(...)} accepts arrays, file paths, level mappings, or attribution bundles, and \texttt{add\_overlay(...)} accepts arrays, NetCDF variables, or overlay bundles, so any producer matching these interfaces can drive the viewer.
During ingestion, each array is cast to \texttt{float32}, validated as a two-dimensional grid, then stored by method, frame, and layer.

From this store, one versioned viewer payload is built.
Each grid is normalized, quantized to an unsigned-byte texture, and base64 encoded: signed attribution maps are scaled symmetrically around zero so positive and negative contributions remain distinguishable after quantization, while sequential overlays use a global minimum and maximum across all frames.
Every texture carries target, timestamp, color map, layer label, and content-hash metadata, where the hash lets the frontend detect unchanged data and skip redundant reloading.
This payload is then synchronized either with the widget or with the browser.
A common \texttt{AttributionViewer} then decodes the payload into typed dense grids and uploads the selected attribution and overlay frames as WebGL textures. In map mode these textures are drawn through MapLibre custom layers, while globe mode renders them with Three.js on the globe geometry. In both cases shader programs sample the encoded grids, apply color maps, opacity, smoothing, contours, and frame blending.

\section{Result, Target, and Overlay Formats}
\label{appendix:result-format}
\begin{figure*}[tbp]
  \centering
  \includegraphics[
    width=0.94\textwidth,
    height=3.7in,
    keepaspectratio
  ]{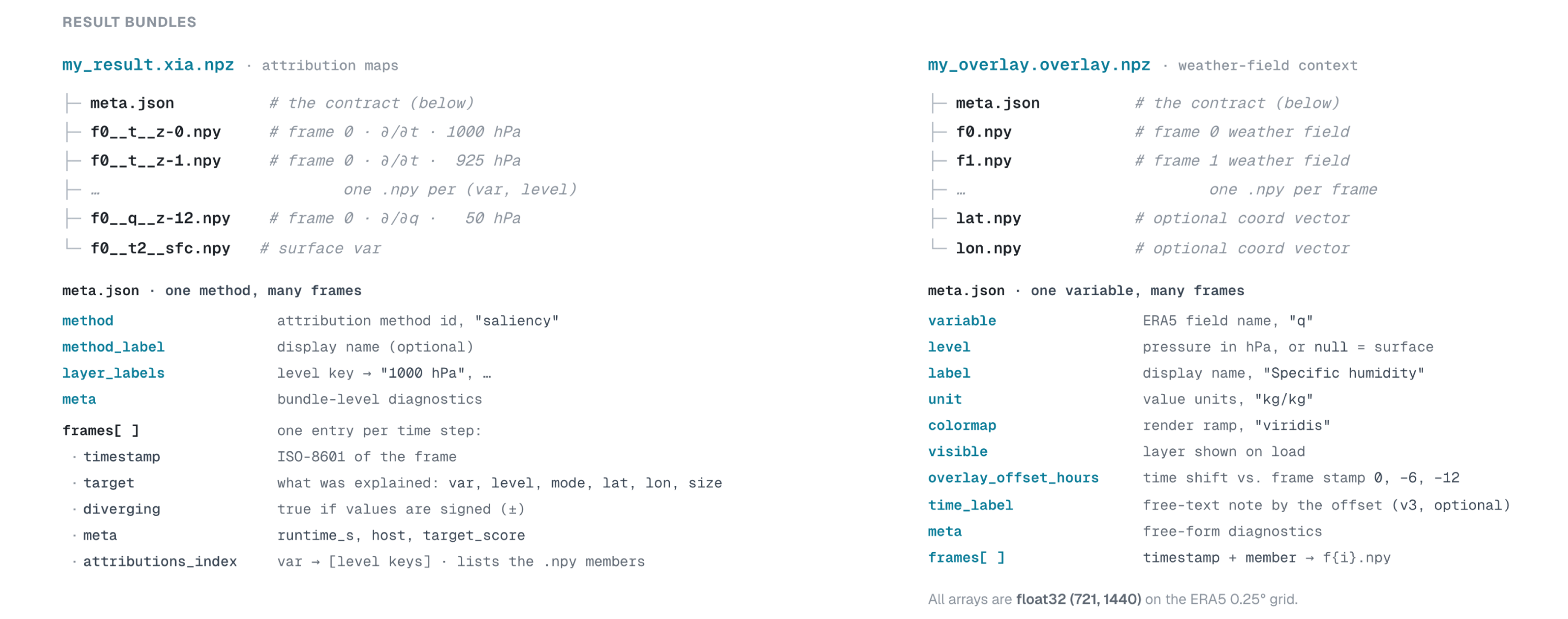}
  \caption{GeoXplain result bundles. Attribution bundles store a manifest plus one array per frame, input variable, and layer. Overlay bundles store one weather field over time with display metadata and optional coordinate vectors.}
  \label{fig:result_bundles}
\end{figure*}

The boundary between \texttt{geoxplain} and a computation backend is a data format rather than a shared Aurora dependency.
The viewer accepts objects that satisfy the attribution and overlay interfaces, while the Aurora adapter is the producer of those objects.

The target specification defines the scalar forecast output to explain: output variable, optional pressure level, timestamp, and spatial mode.
Point targets resolve to the nearest grid cell.
Box targets store a center and full latitude--longitude extent and are evaluated as the mean over that box.

Attribution bundles are stored in-memory or on-disk via \texttt{.xai.npz} files.
Each bundle contains one method and one or more frames. Each frame contains its target, timestamp, divergence flag, metadata, and attribution grids organized as \texttt{attributions[input\_var][level\_id]}.
Surface layers use \texttt{sfc}; atmospheric layers use viewer-level keys such as \texttt{z-N}, with display labels such as \texttt{850\,hPa} supplied separately.
As shown in Fig.~\ref{fig:result_bundles}, the saved archive contains a \texttt{meta.json} manifest and one \texttt{float32} array per frame, input variable, and layer.

Overlay bundles are stored separately as \texttt{.overlay.npz} files.
They describe one physical field through variable, optional level, timestamps, unit, color map, visibility, and one \texttt{float32} grid per frame.
Remote execution uses JSON for requests and status updates, while completed attribution or overlay payloads are returned as \texttt{msgpack} and immediately reconstructed.

\section{Remote Execution Data Flow}
\label{appendix:aurora-data-flow}
\begin{figure*}[tbp]
  \centering
  \includegraphics[
    width=0.94\textwidth,
    height=3.7in,
    keepaspectratio
  ]{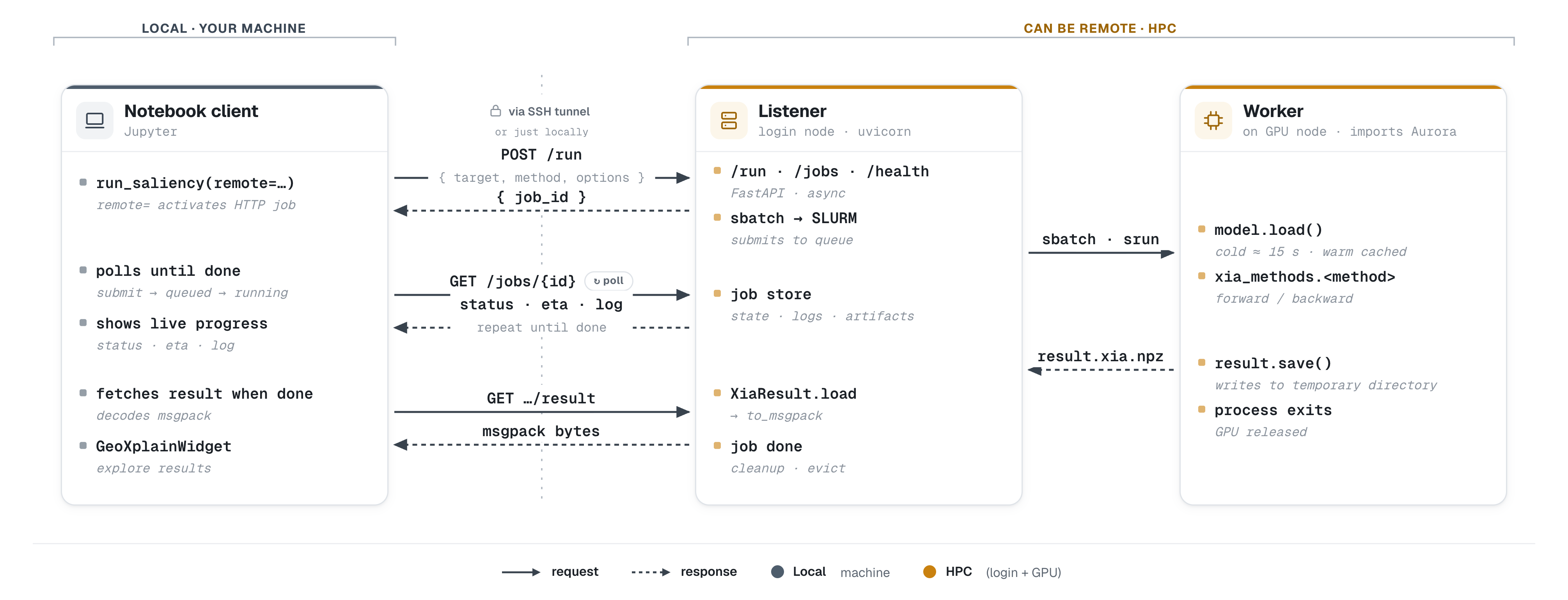}
  \caption{Remote execution in the Aurora adapter. A notebook client submits an HTTP job, polls the listener for progress and logs, and fetches packed result bytes. On HPC systems, the listener can hand the computation to a GPU worker through SLURM while preserving the same client-side result object.}
  \label{fig:architecture}
\end{figure*}

A call such as \texttt{run\_ig(target, input=[...], remote=...)} supplies the target specification, method, and input variables.
Omitting \texttt{remote} runs in the current GPU process; supplying a URL sends the same request to a listener, commonly reached through an SSH tunnel on HPC systems.
The listener can run work in-process, forward to a persistent GPU worker, or submit one SLURM job per request, as summarized in Fig.~\ref{fig:architecture}.

The HTTP API is intentionally narrow.
\texttt{POST /run} submits one attribution target, \texttt{POST /run\_batch} submits multiple target timestamps, and \texttt{POST /overlay} requests raw ERA5 overlay fields, each of which returns a \texttt{job\_id}.
The client then polls \texttt{GET /jobs/\{id\}} for queue state, progress, ETA, log tail, and the result URL.
When the job is complete, \texttt{GET /jobs/\{id\}/result} returns an attribution or overlay result bundle, and \texttt{GET /health} reports backend mode and queue state.

After dispatch, the adapter interprets the target timestamp as Aurora's second input time step \(t_1\) and loads the matching case with \(t_0=t_1-6\) hours.
The requested inputs are split into atmospheric and surface tensors, and the target resolver turns the target specification into a differentiable scalar over Aurora's prediction: a nearest-cell point value or a box mean at the requested surface or pressure level.

Method runners return attribution maps keyed by input variable.
Saliency and Integrated Gradients split atmospheric output into per-level grids and store surface output under \texttt{sfc}. RISE and ViT-CX perturb atmospheric columns horizontally and therefore store one level-collapsed atmospheric map.
The adapter wraps the maps, labels, divergence flags, runtime metadata, and optional target scores into an attribution result bundle. Batch time frames and rollouts add further frames.

\end{document}